\documentclass{article}

\usepackage{arxiv}

\usepackage[utf8]{inputenc}
\usepackage[T1]{fontenc}    
\usepackage{hyperref}       
\usepackage{url}            
\usepackage{booktabs}       
\usepackage{amsfonts}       
\usepackage{nicefrac}       
\usepackage{microtype}      
\usepackage{graphicx}

\title{Efficient direct slicing of dilated and eroded 3D models for additive manufacturing}

\author{
  Sylvain Lefebvre \\
  Inria Nancy Grand-Est\\
  LORIA, Université de Lorraine\\
  \texttt{sylvain.lefebvre@inria.fr} \\
}

\renewcommand{\undertitle}{Technical Report}

\begin{document}
\maketitle

\begin{abstract}
In the context of additive manufacturing we present a novel technique for direct 
slicing of a dilated or eroded volume, where the input volume boundary is a triangle mesh.
Rather than computing a 3D model of the boundary of the dilated or eroded volume, our technique directly 
produces its slices. This leads to a computationally and memory efficient algorithm, 
which is embarrassingly parallel. Contours can be extracted under an arbitrary chord error, 
non-uniform dilation or erosion are also possible. Finally, the scheme is simple and 
robust to implement.
\end{abstract}

\section{Introduction}
\label{sec:intro}

Most additive manufacturing (AM) process planning tasks start with the \textit{slicing} operation~\cite{Attene:2018:DRP}, where a model is decomposed into a set of planar slices. Each layer is then transformed into a set of instructions describing how the AM machine will solidify an uncured material (resin, powder, filament) into a solid slab reproducing the slice geometry.

Slicing is typically performed onto a tessellated version of the geometry,
often provided as a triangle mesh $\mathcal{M}$ under the STL file format. This mesh is expected to properly define the boundary of an inner volume $V$: it has to unambiguously separate the inner volume from the outside. 
This is typically defined in the winding number sense~\cite{jacobson13}, using facet orientations to define outside-to-inside and inside-to-outside interfaces. 
By convention, we consider that faces with counter clock wise (CCW) orientation define an interface from outside to inside (entering), while faces with clock wise orientation define an interface from inside to outside (exiting).  
This definition is convenient as it naturally captures holes (cavities), and is robust to self-intersections. 

Slicing computes the intersection of a set of slicing planes with the geometry, forming a set of \textit{slice contours}: general 2D polygons representing the intersection area of each slice. Note that other definitions of slicing are possible, where the slice contour is not the result of an intersection~\cite{Alexa:2017:ODS}. However, in this work we follow the typical point of view of intersecting the model with a plane at each slice height. 

When slicing an oriented 3D triangle mesh, the contour orientation is trivially carried from the triangles to the polygon outline segments. Thus, the obtained polygons also obey the winding number rule. Their orientations hence define solid and empty areas (holes). These contours can for instance directly be fed to the OpenGL tessellator to obtain a valid area triangulation through the winding rules, with the added benefit of numerical robustness~\cite{McMains:2000:thesis}. The contours of the triangulated areas form the final, cleaned, slice contours~\cite{McMains:2000:thesis}.

While most slicers input triangular meshes, tessellation can be avoided by defining \textit{direct} slicing procedures~\cite{Jamieson:1995:DSO}, that manipulate the input geometry under its original representation. For instance, slicing a sphere equation leads to disks equations, slicing a cylinder (away from its extremities) leads to ellipses \cite{Rosen:2007}. In such cases, the produced contours remain under an analytical representation (equations) and thus suffer no loss of precision. 

While slicing is a relatively simple process to implement, the process planner often needs to perform more advanced manipulations on the initial 3D geometry. One operation in particular is highly desirable for process planning: the ability to manipulate eroded and dilated versions of the input volumes. Dilated volumes $V^\uparrow_r$ are defined by the Minkowski sum of a ball $\mathcal{B}_r$ of radius $r$ with the input volume $V$. While the eroded volume $V^\downarrow_r$ is obtained through the dilation of the complement of $V$. In the remainder of this document, for the sake of clarity we focus explanations on obtaining the dilated volume $V^\uparrow_r$. 

Naturally, this problem has attracted a lot of attention in the research community, both in a general setting and in the specific context of AM. Most techniques attempt to reconstruct a mesh $\mathcal{M}^\uparrow_r$ that captures the boundary of $V^\uparrow_r$. A variety of approaches have been proposed, typically using discrete representations (ray-rep~\cite{chiu1998using,wang2013gpubased}, distance fields~\cite{frisken2000adaptively}, zonotopes~\cite{martinez2015chained}) or directly manipulating triangle meshes~\cite{forsyth1995shelling,campen2010polygonal}. Most often these approaches do not compute the offset with a perfect ball, but rather with a discretized version of $\mathcal{B}_r$ (tessellation, voxelization). 
Please refer to Attene et al.~\cite{Attene:2018:DRP} for an overview in the context of AM. Once a mesh is obtained for $V^\uparrow_r$, it can be sliced alongside the input model.

Interestingly, the 3D model of $V^\uparrow_r$ is rarely used directly, only its slices are necessary to define regions of interest within the input model slices.
Park~\cite{park2005hollowing} proposed an approximate method based on this observation, where the slices of a dilated model are obtained from the slice contours of the input model. Each original contour is dilated by a ball (or equivalently, a circle is swept along the curve), which results in a 3D torus-like 'inflated' curve. 
The union of these 3D shapes with the original volume give an approximation of $V^\uparrow_r$. Rather than performing a 3D union, the union can be performed in the slices. Each inflated curve, when intersected by a slice below or above, produces two contours. The contours may be obtained by dilating the center curve by a radius which depends on the vertical distance between the slice and the curve. This implies that the slices of the approximation of $V^\uparrow_r$ are easily computed using only 2D contouring operations.
The main drawback of the technique is that it may miss features which are smaller than the layer spacing used for the dilation. When dilating (or eroding) with a relatively large $r$, these features may have a significant impact on the result. Another difficult case occurs along near-flat surfaces (tops/bottoms) where the curves may become spaced by more that $r$ leading to artifacts.

In this paper we propose an approach that also operates directly in the slices. We define a direct slicing procedure for the \textit{exact} offset volume $V^\uparrow_r$: there is no approximation compared with slicing the boundary mesh $\mathcal{M}^\uparrow_r$. On the contrary our technique affords for a direct control of the chord error with respect to $\mathcal{B}_r$, and an exact representation may be extracted. Our approach results in a highly efficient, robust and scalable parallel algorithm. 

\section{Method}

We seek to slice the dilated volume $V^\uparrow_r = V \oplus \mathcal{B}_r$, where $\mathcal{B}_r$ is a ball of radius $r$. $V^\downarrow_r$ is easily obtained by a similar process, which we detail later for the sake of clarity.

\begin{figure}[tb]
\centering
\includegraphics[width=\linewidth]{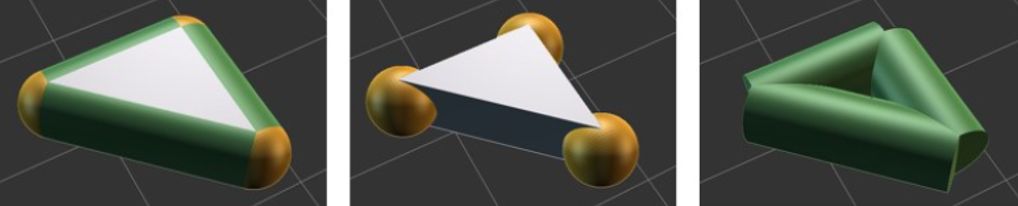}
\caption{Left: A dilated triangle $t_i \oplus \mathcal{B}_r$  and its decomposition. Middle: center prism and spheres. Right: cylinders.}
\label{fig:trioffs}
\end{figure}

We now consider generating a single slice of $V^\uparrow_r$. The slice $j$ is a plane $\mathcal{P}_j$ in space. Let us denote by $t_i$ the i-th triangle of $\mathcal{M}$. A key observation is that a dilated triangle $t_i \oplus B_r$ can be decomposed into three cylinders, three spheres, and a center prism (see Figure \ref{fig:trioffs}).

We slice each of these primitive with $\mathcal{P}_j$; that is, for each triangle $t_i$ we directly slice the three cylinders, spheres and the prism\footnote{as spheres and cylinders are shared by neighboring triangles we can avoid redundant computations} with $\mathcal{P}_j$. We thus obtain a set of CCW contours. 
Direct slicing of simple primitives can be done with high-performance specialized codes: slicing of a sphere produces a circle, direct slicing of a capped cylinder produces ellipses or clipped ellipses (including the rectangle case), direct slicing of a prism produces a triangle or a quad. These contours may be obtained as tessellated contours under a given chord error, or may be retained as exact contour primitives (circular and elliptic arcs, segments).
The triangles may be processed in parallel, and only those triangles potentially intersecting the slice $\mathcal{P}_j$ are considered.

\paragraph{Finding the set of slices intersecting $t_i \oplus B_r$:} Given a triangle $t_i$, the set of slices potentially intersecting $t_i \oplus \mathcal{B}_r$ is easily computed. Assuming the slices are along the $z$ axis, let us denote the $z$ interval covered by the triangle as $[z_i^{min},z_i^{max}]$ (computed from the triangle vertices $z$ coordinates min/max). 
The interval covered by $t_i \oplus \mathcal{B}_r$ is $[z_i^{min}-r,z_i^{max}+r]$.
Given a slicing height sequence $h_0,...h_n$, each specifying the $z$ location of a slicing plane $\mathcal{P}_j$, we efficiently retrieve the indices of the first and last slice intersecting  $t_i \oplus \mathcal{B}_r$. In case of a uniform slicing thickness $\tau$ this is a direct computation with a rounding, e.g. $j_{first} = \lfloor \frac{z_i^{min}-r}{\tau} \rfloor$. In case of adaptive slicing we perform a logarithmic bisection search in the sequence of slice heights, locating the first slice just above $z_i^{min}-r$ (and similarly for the last slice).

When processing in parallel, we launch $K$ threads that visit all triangles simultaneously. Each thread takes a next triangle, determines which slices it impacts, generates the dilated triangle contours for each slice, and stores them in the slice data structure.
 Once all triangles have been visited by all threads, the final contouring is performed in all slices, again in parallel.
As storing the contours may lead to a high memory requirement, this process may be performed for a subset of slices only. For instance, in a first pass, the contours are produced only for the $N$ first slices and discarded for others.
Once done with the $N$ first slices, we proceed to the next $N$ slices and start over. This allows to fit computations within a constrained memory budget, with minimal computational
redundancy (visiting the triangles being very fast, the cost is in generating and storing the contours).

\paragraph{Producing the final contours of $V^\uparrow_r$:}
The CCW contours obtained from the triangles are simply added to the list of contours obtained by slicing $\mathcal{M}$ with $\mathcal{P}_j$. 
The winding number contouring directly yields a slice of $V^\uparrow_r$. 

Similarly, a slice of $V^\downarrow_r$ is obtained by reversing the contours from the dilated triangles to be CW, before extracting the final contour with
the winding number rule. This follows from the definition of the erosion as "the complement of the dilation of the complement" ($V^\downarrow_r = \overline{ \overline{\mathcal{M}} \oplus \mathcal{B}_r}$): 
the complement first reverses the contour of $\mathcal{M}$ (solid areas become holes, while holes become solid). We then add the CCW contours obtained from the triangles to dilate,
and then reverse every contour again. Thus, in the end we only have to revert the contours from the dilated triangles.

Note that on AM processes that take images as input (e.g. DLP SLA), the final contour extraction may be skipped. Rasterizing all polygons using
the winding rule in the stencil buffer will yield the final, correct image.

\paragraph{Contour extraction:}
By default we produce tessellated contours for the spheres and cylinders, enforcing a target chord error. Given these, any polygon contouring algorithm with winding number support may be used (e.g. Vatti's algorithm \cite{Vatti:1992:GSP} or the marching intersections \cite{Rocchini:2001} among many choices in the literature). These can be further sped up, noting that the additional contours produced by the spheres/cylinders/prisms are all convex. This allows to skip some computations,  since the number of intersections of a sweep line with convex primitives is always exactly two (excluding exactly aligned vertices).

Exact contours may be obtained by computing proper intersections between circular and elliptic arcs and polygons. Quite remarkably this would produce the contours
of the dilation of $V$ by an exact ball $\mathcal{B}_r$. We are not aware of any other technique able to achieve this.

When computing contours from dilated triangles in parallel, the contours are sorted by triangle id before contouring, ensuring a deterministic result.

\begin{figure}[tb]
	\centering
	\includegraphics[width=0.22\linewidth]{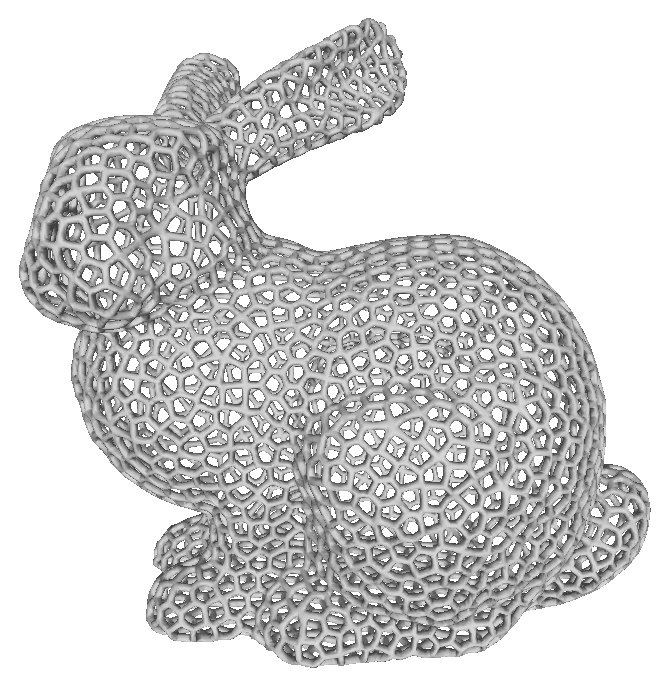}\hfill
	\includegraphics[width=0.7\linewidth]{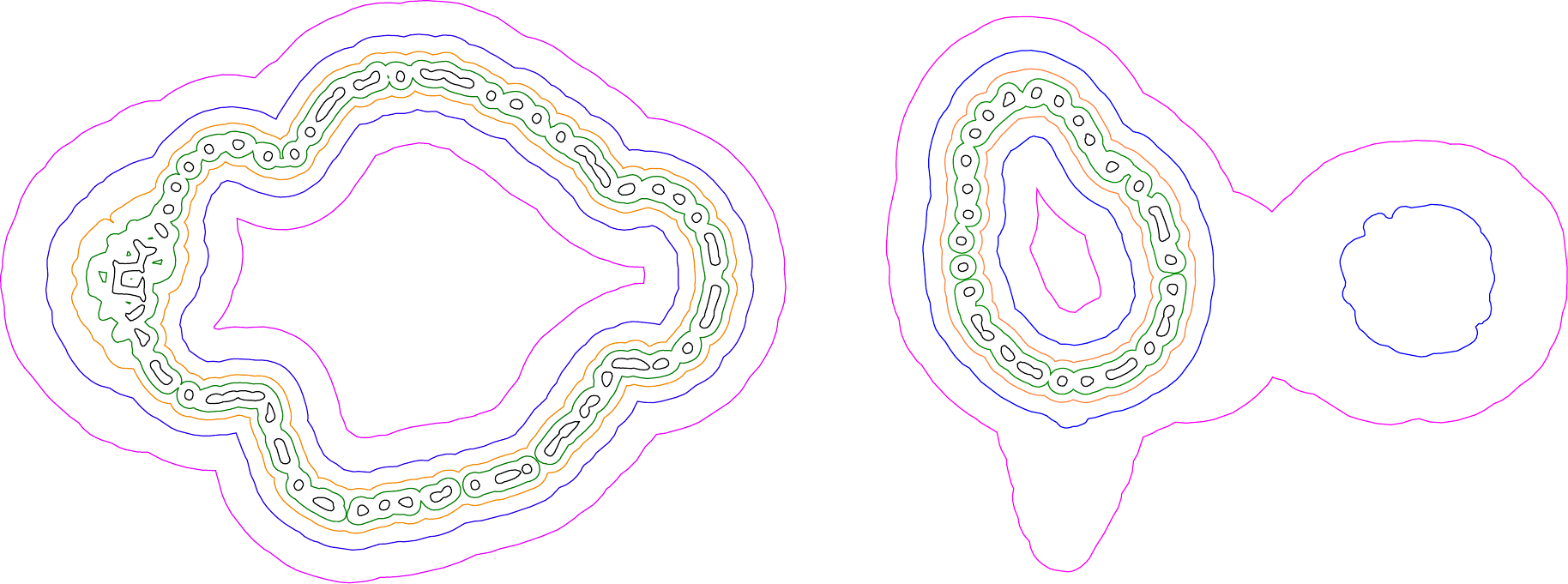}
	\caption{Slices of the Voronoi Bunny model for different dilation radii $r$: original (black), 1 mm (green), 2 mm (orange), 4 mm (blue), 8 mm (purple). The rightmost slice is located through the Bunny's head ; we can see the back of the bunny appearing with large dilation on the right side.
		\textit{Model:} Voronoi Stanford Easter Bunny by Virtox \url{https://www.thingiverse.com/thing:303842}
	}
	\label{fig:results}
\end{figure}

\section{Properties and results}

Figure~\ref{fig:results} shows slices from a complex 3D model with various amounts of dilation. The model
has half a million triangles with intricate geometry, and  would be typically very challenging for
mesh-based algorithms. We slice the model at $40\mu m$, resulting in 2379 slices (before dilation).

\begin{figure}[tb]
	\centering
 	\includegraphics[width=0.45\linewidth]{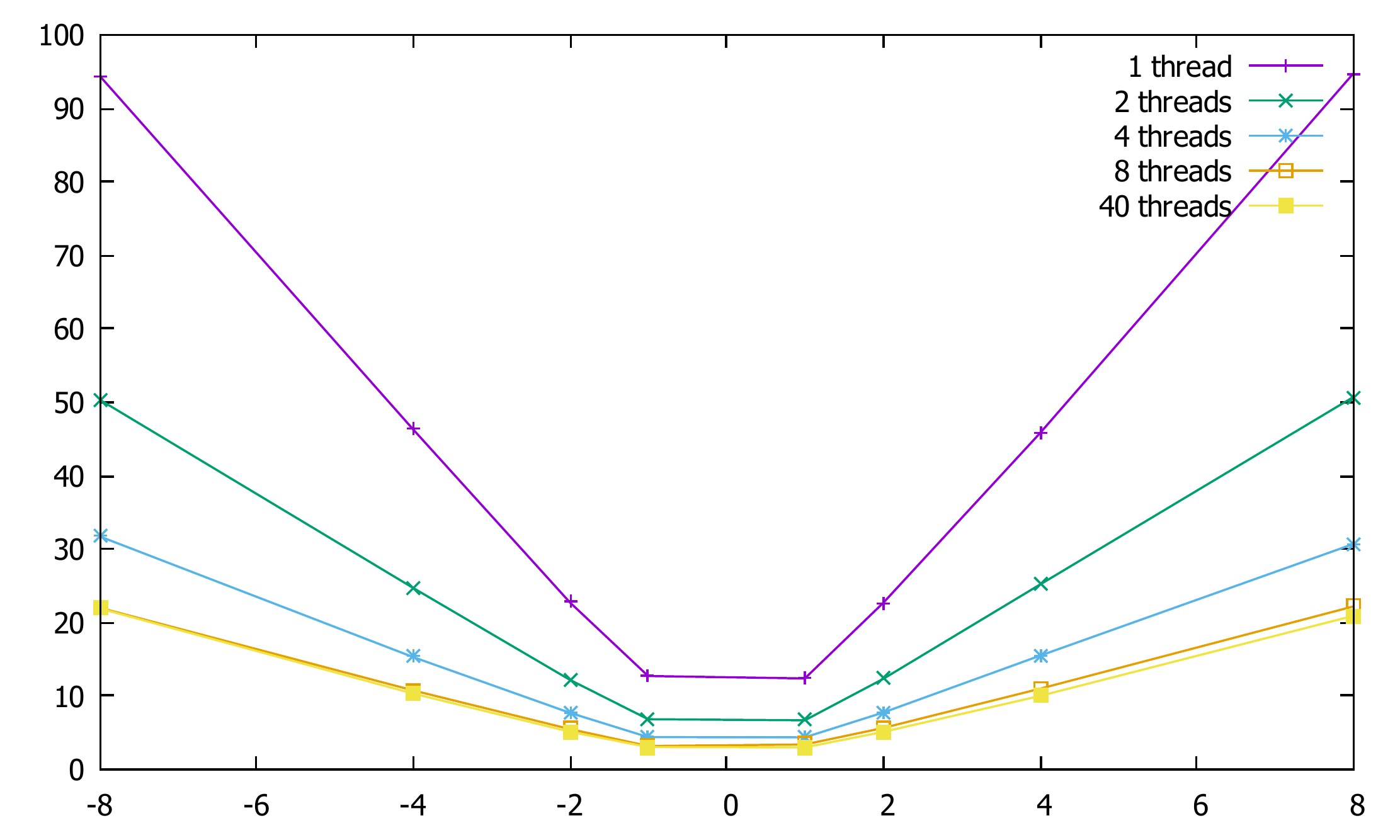}
 	\hspace*{-1cm}\includegraphics[width=0.07\linewidth]{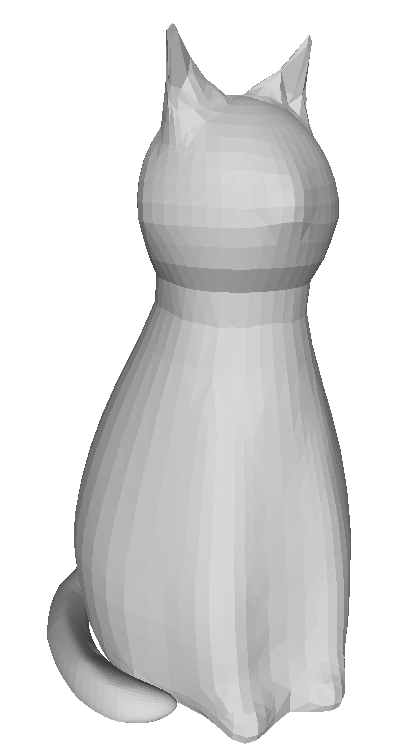} \hfill
 	 \includegraphics[width=0.45\linewidth]{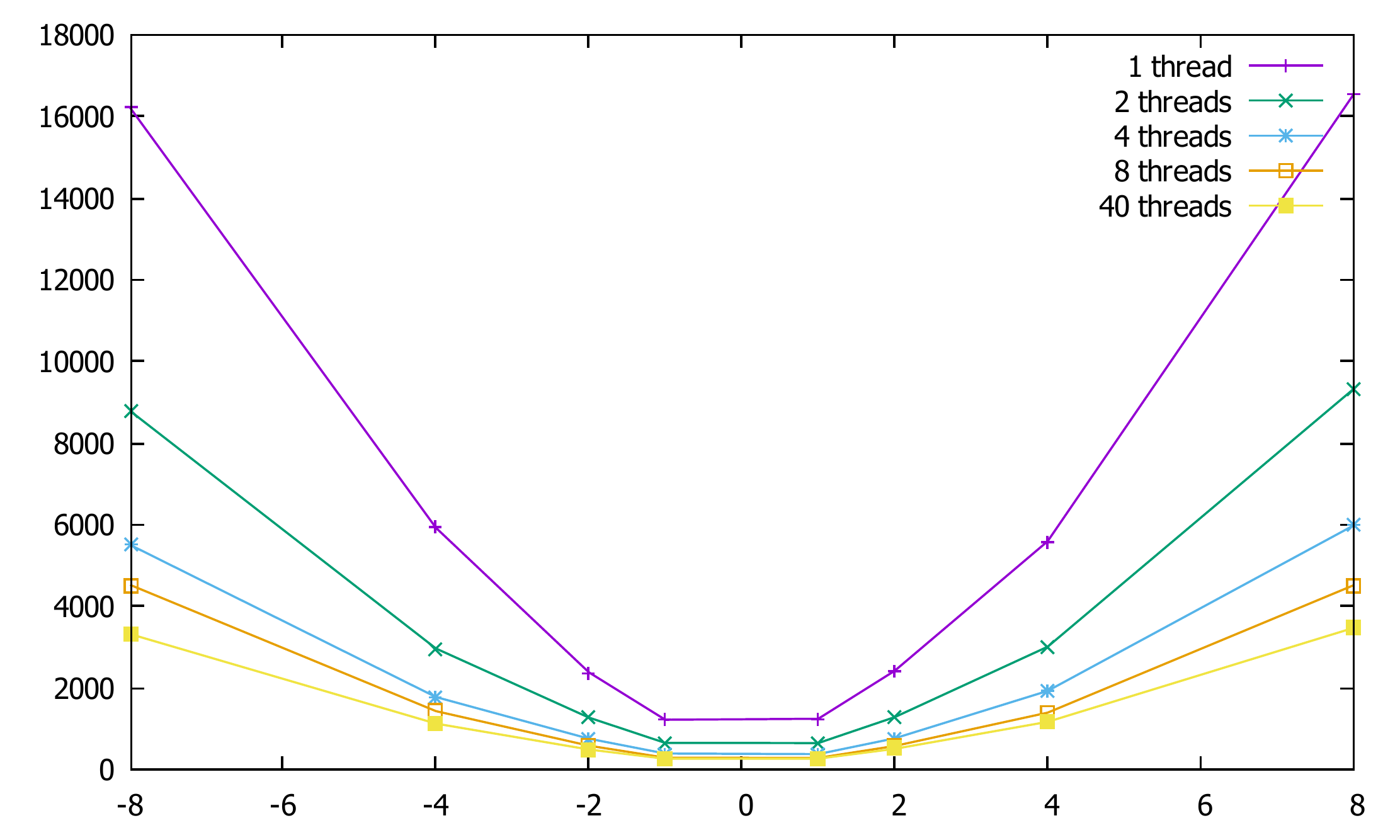}
 	 \hspace*{-1cm}\includegraphics[width=0.1\linewidth]{vbunny.png} \\
 	\hspace{1.7cm} \textbf{Lucy the cat (8K triangles)} \hfill \textbf{Voronoi Stanford Easter Bunny (561K triangles)}
	\caption{Performance graphs for different radii $r$ and number of threads, lower is better. The vertical
	axis is in seconds, the bottom in mm ($r$). A negative $r$ indicates dilation, a positive $r$ an erosion. 
	Each curve is for a different number of threads, from 1 to 40.
\textit{Models:} lucy the cat by Mere \url{https://www.thingiverse.com/thing:24255}, Voronoi Stanford Easter Bunny by Virtox \url{https://www.thingiverse.com/thing:303842} }
	\label{fig:perfs}
\end{figure}

The complexity of computing the contour depends essentially on the number of triangles possibly intersecting a slice. This is a function of triangles per slice (tessellation factor of the input) as well as $r$ (the greater $r$ the more slices are impacted by each dilated triangle). 
Figure~\ref{fig:perfs} reveals the performance behavior as a function of the radius $r$ and the number of threads. Measurements are conducted on an Intel i7-4790K CPU, four physical cores, with 32GB of memory. As can be seen performance increases linearly with the number of threads --- the algorithm is embarrassingly parallel --- until the number of physical cores (four) is exceeded. After this point, smaller improvements are still observed (these occur thanks to hyper-threading and improved scheduling/latency hiding when having many threads).
The relationship between $r$ and performance is more subtle. As $r$ increases additional triangles impact each slice, thus increasing the cost. In the case of the Lucy cat model the increase is near linear (this is the case with most smooth models), while it is quadratic with the Bunny model. This depends greatly on the input geometry complexity.

An advantage of our technique is that slices may be computed independently, in any order, or in parallel, every time yielding the same result with consistent performance. In particular, consider the case of previewing a single slice of a dilated version of $V$. With a standard technique, the entire boundary mesh of  $V^\uparrow_r$ has to be computed, before a single slice is extracted. With our technique, the computations are performed only for the slice, and most triangles are completely ignored.

The limiting factor in terms of performance is the contouring algorithm, as the number of contours produced from the dilated triangles can become quite large per-slice.
We found it best to perform contouring progressively (everytime $k$ contours are stored a new contour is extracted with the winding rules, which accumulates
all small contours in a partial result) or equivalently having a divide-and-conquer strategy at the end, recursively performing contouring on half-subsets of all the contours
(the subset not being only two contours, but typically a larger number, determined experimentally).
Accumulating all the contours before computing the final one with the winding rule usually leads to significantly slower results.

Finally, it is worth noting that our algorithm trivially generalizes to varying dilation/erosion, where the ball radius $r$ varies in the vertices
of the input mesh (and is linearly interpolated along edges and triangles). The only change is that dilated triangles produce (truncated) cones instead of 
cylinders, and the center prism faces become sloped. The radius may even become negative, effectively resulting in a mixed dilation-erosion operator.

However, this algorithm cannot be used to perform opening or closing morphological operations, since the starting point has to be a triangle
mesh, and we cannot extract the triangle mesh of the intermediate step. 

\newpage

\bibliographystyle{unsrt}  
\bibliography{references}  

\end{document}